\documentclass[conference]{IEEEtran}
\usepackage{graphicx}
\usepackage{gensymb}
\usepackage[export]{adjustbox}
\usepackage[font=small,labelfont=bf]{caption}
\ifCLASSINFOpdf
\else
\fi
\begin{document}
\title{Current Mode Neuron for the Memristor based synapse}
\author{\IEEEauthorblockN{Harshit Roy}
\IEEEauthorblockA{Electronics and Electrical Communication Engineering\\
Indian Institute of Technology\\
Kharagpur, West Bengal\\
Email: harshit.roy@iitkgp.ac.in}
\and
\IEEEauthorblockN{Mrigank Sharad}
\IEEEauthorblockA{Electronics and Electrical Communication Engineering\\
Indian Institute of Technology\\
Kharagpur, West Bengal\\
Email: mriganksh@gmail.com
}}

\maketitle

\begin{abstract}
Due to many limitations of Von Neumann architecture such as speed, memory
bandwidth, efficiency of global interconnects and increase in the application of
artificial neural network, researchers have been pushed to look into alternative architectures
such as Neuromorphic computing system. Memristors (memristive crossbar memory
RCM) are used as synapses due to its high packing density and energy efficiency and
CMOS blocks as neurons. The increase in the terminal resistance of the RCM can
degrade its energy efficiency and bandwidth. A more energy efficient current mode
neuron has been proposed in this paper which can operate at lower voltages as
compared to conventional voltage mode neuron circuit.
\end{abstract}


\IEEEpeerreviewmaketitle

\section{Introduction}
The increasing demand for computation and decreasing size of transistors have led to the integration of multiple cores and memory on a single chip. The number of buses as well as the resistance per unit length of metal wire on the chip has increased\cite{4378787}. These factors have put restriction on the speed of computation. In recent years, the popularity of the Neural Network has grown due to the increase of its application in image processing, etc. Computation (such as dot product) at the sensor level can reduce the number of buses, high-speed data converters and interface circuits.
The memristor crossbar array is used to build an analog neural network which can compute dot products which are widely used in the image processing application. One of the important modules of the neural network is the activation function like sigmoid function which is used in clustering and pattern recognition problems. \newline
Mostly the sigmoid function is implemented in voltage mode\cite{6706772}\cite{7858397}, but in this paper, we have proposed a more effcient current mode neuron for memristor-based synapse and compared it with the voltage mode neuron in terms of power, bandwidth, input impedance and robustness.
With current mode processing, one can work with lower values of voltages, as has been shown in previous work\cite{Bashirullah:2003:CSD:944103.944114} . This presents the possibility of achieving significant reduction in power consumption.\newline
This paper is organized into the following sections: Section II introduces us to the memristor crossbar array, the relation between input current or voltage with the output current and the effect of parasitic and terminal resistance on power, bandwidth and accuracy. Section III contains voltage mode CMOS sigmoid activation function circuit for meristors with its simulation result and compares it with other sigmoid function circuits in voltage mode. Section IV contains the proposed current mode sigmoid activation function circuit for memristors with the detailed simulation results and compares the low input impedance stage (current to voltage converter) used in this paper with another current to voltage stage. Section V compares the voltage mode circuit with the proposed current mode circuit.

 



\section{RCM Computation}
A memristor,as proposed by Prof. Leon Chua in 1971, is a passive two-terminal fundamental element, in addition to the existing resistor, capacitor, and inductor. It follows a non-linear relation between terminal voltage \( \mathit{v(t)} \) and resultant current \( \mathit{i(t)} \), given by \cite{6706772}
\begin{equation}
    i(t)=G_m(x).v(t) 
\end{equation}
\begin{equation}
    \frac{\partial x}{\partial t} = i
\end{equation}
where $G_m(x)$ is not a constant but a state dependent trans-conductance.A memristor has two regions- high concentration dopant region of resistance $R_{on}$ and low concentration dopant region of resistance $R_{off}$. The total resistance which is a sum of $R_{on}$ and $R_{off}$ can be changed by applying a voltage bias \cite{Strukov2008}.Hence the net tranconductance is given by 
\begin{equation}
    G_m(x) = G_{moff}.(1-x)+G_{mon}.x 
\end{equation}
A specific amount of energy (or threshold voltage) is required to change the stage of a memristor, and below the threshold voltage, memristor behaves as a constant transconductance or resistance\cite{6241552}. A memristor crossbar array has very high density, low power consumption, and variable transconductance. These properties make it suitable to be used as synapses in an Artificial Neural Network or dot product operation which is heavily computed in Image processing.\newline A \( \mathit{N*N} \) memristor crossbar array consists of horizontal\( \mathit{(i)} \) and vertical\( \mathit{(j)} \) metal lines which are connected by memristors of conductance \( \mathit{g_{m_{ij}}} \).These metal lines contribute to the parasitics which can be modeled as parasitic resistance $R_p$ between two nodes and parasitic capacitance $C_p$ between a node and ground. In voltage mode (Fig. 1), \( \mathit{V_1,V_2,..,V_i,..,V_N} \) voltages are applied to the corresponding horizontal lines\( \mathit{(i)} \) and resultant currents\( \mathit{(I_j)} \) from the vertical lines\( \mathit{(j)} \) are converted into voltages by using opamp circuits\cite{6241552}\cite{6706772}. The resultant current in a vertical line \( \mathit{(j)} \) is given by
\begin{equation}
I_j = \sum_{i=1}^{n} g_{m_{ij}}.V_i
\end{equation}
\includegraphics[scale=0.5]{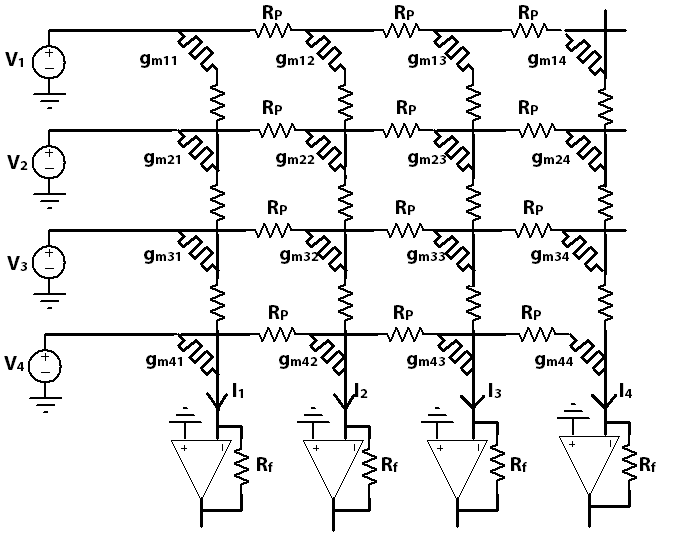}
\captionof{figure}{Voltage Mode Neural Network}
\includegraphics[scale=0.45]{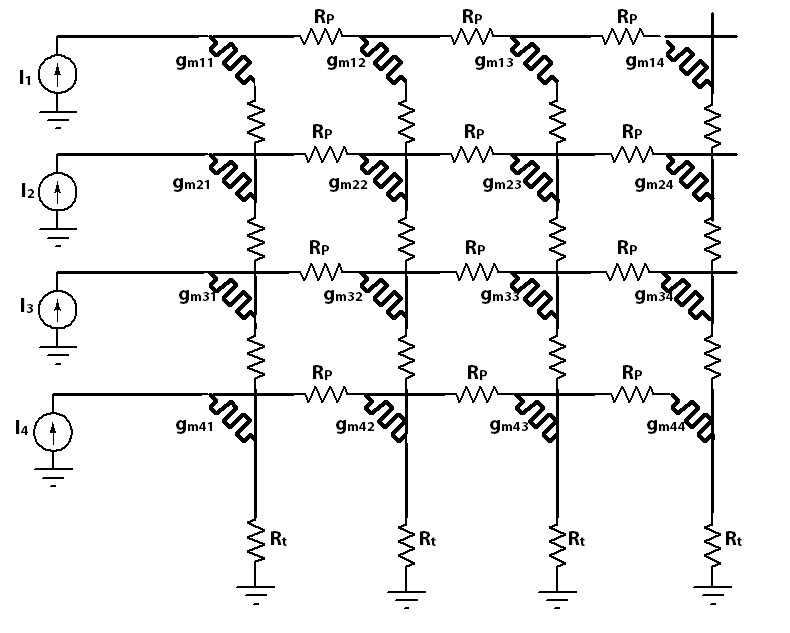}
\captionof{figure}{Current Mode Neural Network}
Similarly, in the current mode (Fig. 2),\( \mathit{I_1,I_2,..,I_i,..,I_N} \) ,input currents are applied to the corresponding horizontal lines\(\mathit{(i)} \) and the resultant current in the vertical line is given by 
 \begin{equation}
I_j= \sum_{1}^{n}I_i.\frac{g_{ji}}{g_{1i}+g_{2i}+..+g_{ji}..+g_{Ni}}
\end{equation}
Equations (4) and (5) are valid provided  that  the  input  impedance (Terminal Resistance $R_t$) of  the  current sensing circuit  is very low.The conductance of the last memristor has to be set in such a way that the sum of the memristor conductance i.e $g_{1i}+g_{2i}+..+g_{ji}..+g_{Ni}$ is same for every $i^{th}$ row.
\subsection{Simulation Results}
The effect of parasitic $R_p$ and terminal resistance $R_t$ on bandwidth, energy, and accuracy of the dot product through numerical analysis performed on MATLAB  are shown in Figure 3. In Figure 3(a), we observe that the bandwidth of the RCM network decreases with an increase in the terminal resistance. Figure 3(b) and Figure 3(d) show the deterioration in the accuracy of the dot product operation with increasing $g_m$ (conductance of memristor) and decreasing $g_t$ (inverse of terminal resistance $R_t$). The effect of increasing terminal resistance on the energy per computation is shown in Figure 3(c). We observe that beyond a critical $R_T$, the energy consumption increases linearly with terminal resistance, i.e., the input impedance of the neuron circuit. Hence, it’s crucial to keep low input impedance ( or terminal resistance $R_t$).
\includegraphics[scale=0.5]{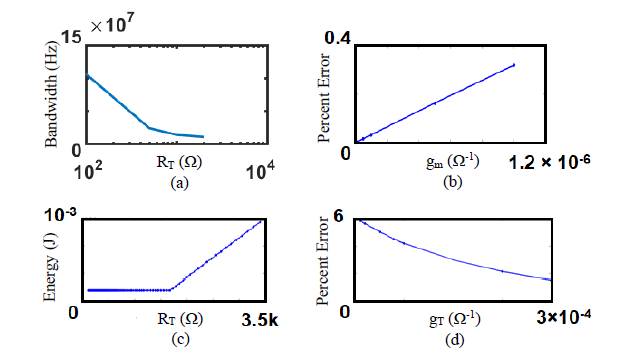}
\captionof{figure}{(a) Bandwidth vs $R_t$ (b) Error in dot product vs $g_m$ (c) Energy Consumption vs Computation (d) Error in dot product vs $g_t$ }
\section{Voltage Mode Sigmoid Neuron}
The circuit consists of a current to voltage converter followed by a single stage differential amplifier in negative feedback and
an inverter in negative feedback. Input impedance is approximately equal to the feedback
resistance R divided by the gain of the opamp. The opamp used here is a compensated
(Compensation capacitor = 200fF and phase margin = 60°) two-stage differential amplifier
where the first stage is differential with current mirror loading and the second stage is a common
source stage. The sigmoid function is given by the below expression
\begin{equation}
   y = \frac{a}{1+e^{b(x-c)}} 
\end{equation}
\includegraphics[scale=0.40]{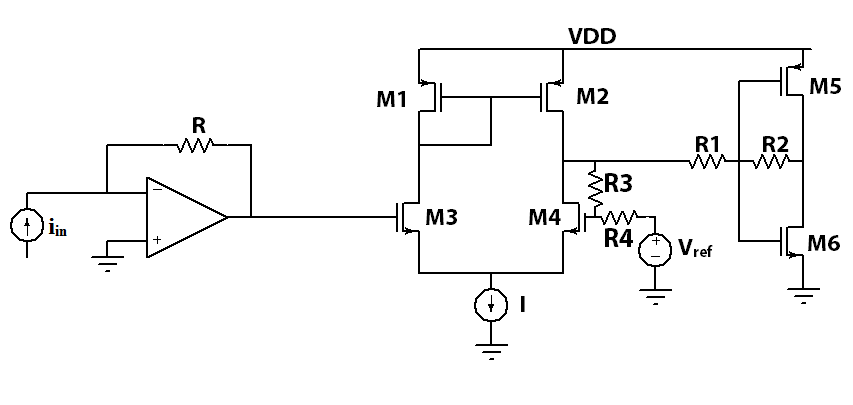}
\captionof{figure}{Voltage Mode Sigmoid Activation Function}
The output of the single stage differential amplifier in negative feedback provides an approximate sigmoid function and gain is controlled by the inverter in negative feedback \cite{59426}.In this case, parameter c depends on reference voltage \( \mathit{(V_{ref})} \) and b depends on the gain of the opamp stage (i.e
\( \mathit{-{R_3}/{R_4})} \)) and gain of the inverter stage (i.e
\( \mathit{-{R_2}/{R_1})} \)) . The inverter stage is used to obtain rail to rail voltage swing as the opamp stage has restricted output voltage swing. The voltage swing decides the value of the parameter ‘a’.
\subsection{Simulation Results}
All the circuit simulation results are performed in Cadence Virtuoso using TSMC 180nm library.In figure 5, dots represent the sigmoid transfer characteristic obtained and it is fitted to an ideal transfer characteristic(as shown). The parameters with 95\% confidence bound are (Table I)
\begin{table}
\centering
\begin{tabular}{ |c|c| } 
\hline
 a & 1.754 (1.723, 1.785) \\ \hline
 b & $-2.13\times10^6 (-2.36\times10^6 ,-1.89\times10^6 )$ \\ \hline
 c & $4.963\times10^{-6} (4.903\times10^{-6} ,5.024\times10^{-6} )$ \\ \hline
 RMSE & 0.06422 \\  \hline
\end{tabular}
\caption{Sigmoid function parameter in Voltage mode}
\end{table}
\includegraphics[scale=0.30]{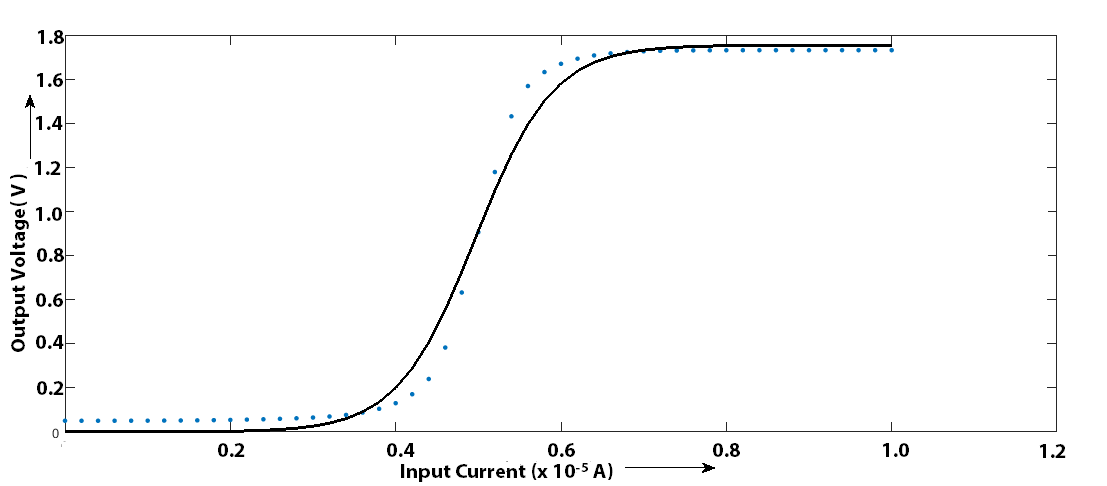}
\captionof{figure}{Sigmoid Tranfer Function obtained from Voltage Mode Neuron}
\includegraphics[scale=0.30]{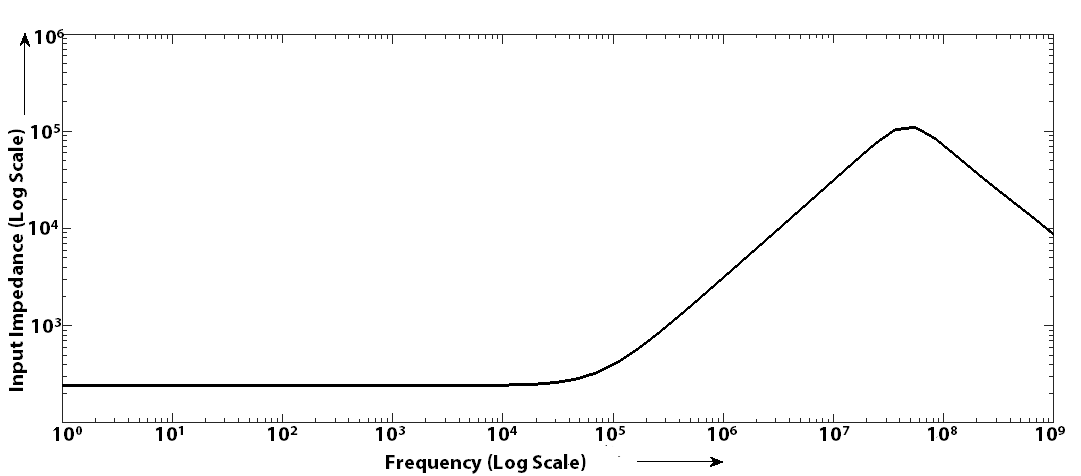}
\captionof{figure}{Input Impedance of the Voltage Mode Neuron}
Voltage mode neuron is operated at lower VDD to improve power consumption. Simulation results show that the deviation in the parameter of the sigmoid function on scaling down the VDD.
\includegraphics[scale=0.30]{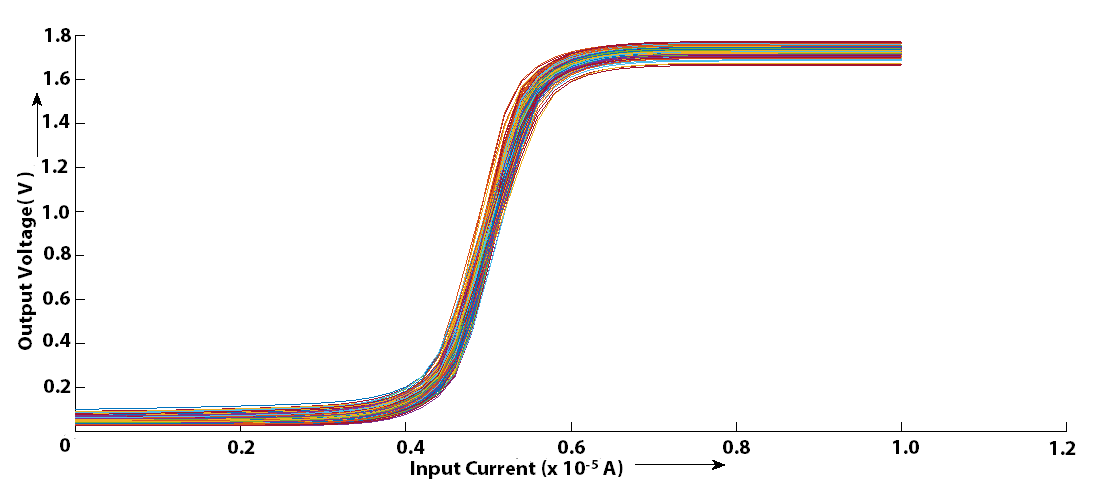}
\captionof{figure}{Monte Carlo Simulation with 200 samples at VDD = 1.8}
The bandwidth of the voltage based neuron is 50Mhz. The maximum current consumption of the circuit is
56$\mu$A (Power = 100.8$\mu$W). On scaling down the voltage, the power consumption of the circuit will decrease
proportionally. Input impedance is 243$\Omega$ with zero at 78Khz.On scaling down the voltage, the
gain of the opamp will decrease slightly which results in a slight increase in the input impedance
to 307$\Omega$ .\newline
In one of the papers, the sigmoid circuit is implemented by the current buffer followed by a sigmoid circuit (Power 160$\mu$W) \cite{DBLP:journals/corr/abs-1808-00737}and in another paper, they have proposed memristor based sigmoid activation function (Power = 0.244mW) \cite{Kafizov2018CMOSMemristiveSA}.The transfer function of sigmoid (reported in this paper) is not symmetric with respect to central horizontal axis \cite{Kafizov2018CMOSMemristiveSA}.

\begin{table}[]{%
\begin{tabular}{|c|c|c|c|c|c|c|}
\hline
VDD & \multicolumn{2}{c|}{a} & \multicolumn{2}{c|}{b} & \multicolumn{2}{c|}{c} \\ \hline
(V) & Mean       & Std \%    & Mean      & Std \%     & Mean      & Std\%      \\ \hline
1.8 & 1.7489     & 1.13\%    & -2.25$\times 10^6$         &  8.4\%          &   4.95$\times 10^{-6}$        &      1.16\%      \\ \hline
1.5 & 1.4979     & 3.52\%    & -2.28$\times 10^6$          &  11.2\%          &    4.94$\times 10^{-6} $      &     1.87\%       \\ \hline
1.0 & 0.9525     & 9.08\%    & -1.44$\times 10^6$          &   13.32\%         &   4.72$\times 10^{-6} $       &    5.96\%        \\ \hline
\end{tabular}%
}
\caption{The variation of the sigmoid function parameters (Monte Carlo Simulation with 200 samples) on scaling VDD}
\label{}
\end{table}
\section{Current Mode Sigmoid Neuron}
The circuit consists of a low input impedance stage followed by a differential stage (Fig.8). The input
impedance is given by the following expression.
\begin{equation}
     R_{in} = \frac{r_{ds7}}{1+g_{m7}.r_{ds7}.A } 
\end{equation}
Where $g_{m7}$ and is the transconductances of $M_7$, $r_{ds7}$ is the small signal resistance of $M_7$, and $A$ is the gain of the opamp. Here opamp has a single stage differential amplifier with a
diode connected NMOS load.
Parameter b of the sigmoid function depends on the transconductance of $M_{3/4}$ and $1/g_m$ of $M_{6/7}$ .
$I_{BIAS2}$ of the right current mirror and $I_{BIAS3}$ decides the value of parameters ‘c’ and ‘a’ respectively.
$C_c$ is the compensation capacitor and $R_c$ is the resistance.
\newline
\includegraphics[scale=0.30]{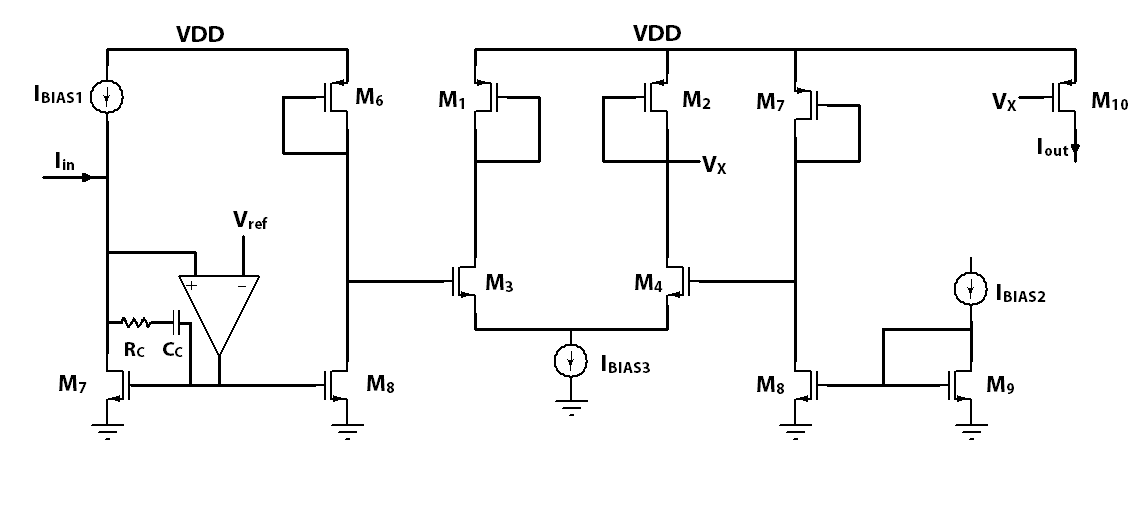}
\captionof{figure}{Current Mode Sigmoid Neuron}
\subsection{Simulation Results}
The value of $C_c$ and $R_c$ is 100fF and 10K$\Omega$ respectively to have 63$\degree$ Phase Margin.
\begin{table}
\centering
\begin{tabular}{ |c|c| } 
 \hline
 a & $4.917\times10^{-6} (4.913\times10^{-6} , 4.92\times10^{-6} )$\\    \hline
 b & $-2\times10^6 (-2.01\times10^6 , -1.99\times10^6 )$ \\    \hline
 c & $2.618\times10^{-6} (2.615\times10^{-6} , 2.62\times10^{-6} )$ \\ \hline
 RMSE & $8.506\times10^{-9}$ \\\hline
\end{tabular}
\caption{Sigmoid function Parameter with 95\% confidence bound}
\end{table}
\includegraphics[scale=0.30]{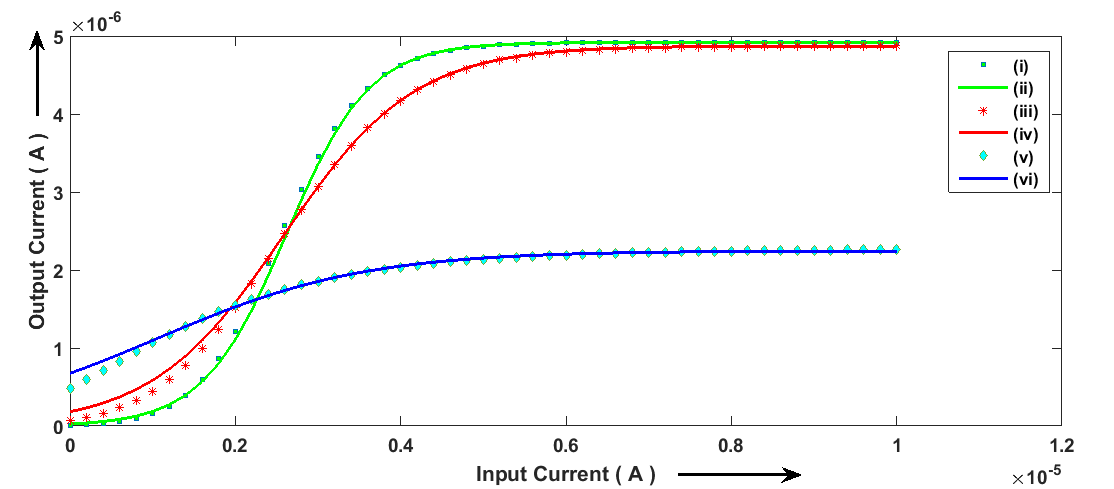}
\captionof{figure}{Sigmoid Transfer Function obtained from Current Mode Neuron }

\begin{table}[]
\centering
\begin{tabular}{|c|c|c|c|}
\hline
VDD & $Z_{in}(ω_{ZERO})$ & Bandwidth & Power \\ \hline
(in V) & $\Omega$(Khz) & (Mhz) & $\mu$ W \\ \hline
 1.8 & 200(144) & 6.25 & 40.5 \\ \hline
 1.5 & 126(130) & 5.2 & 33.75 \\ \hline
 1.0 & 274(266) & 10 & 12.5 \\ \hline
\end{tabular}
\caption{Input Impedance,Power and Bandwidth Variation on scaling voltage}
\label{my-label}
\end{table}
\includegraphics[scale=0.30]{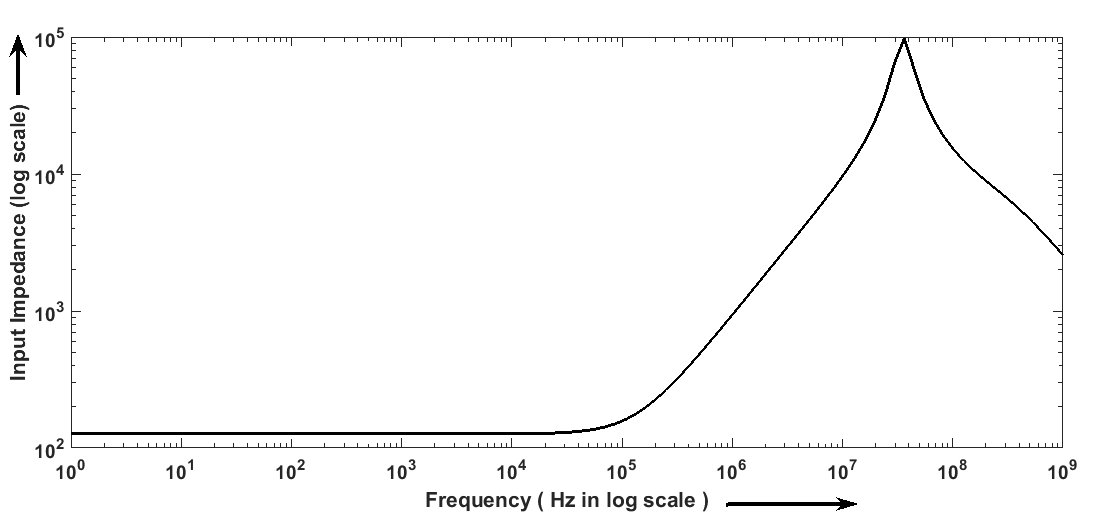}
\captionof{figure}{Input Impedance Of Current Mode Neuron}
\includegraphics[scale=0.30]{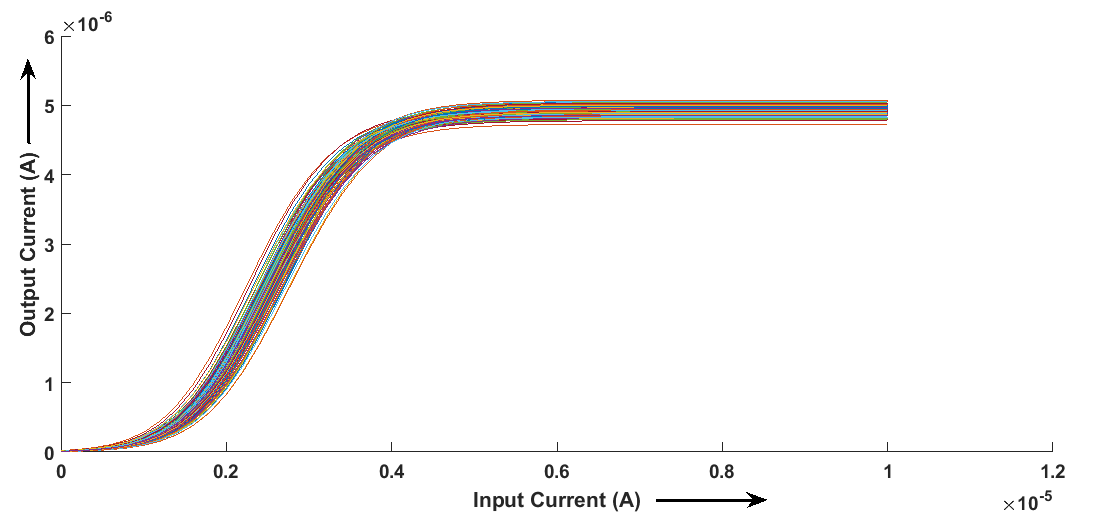}
\captionof{figure}{Monte Carlo Simulation with 200 samples at VDD = 1.8 }
\begin{table}[]{%
\begin{tabular}{|c|c|c|c|c|c|c|}
\hline
Vdd & \multicolumn{2}{c|}{a} & \multicolumn{2}{c|}{b} & \multicolumn{2}{c|}{c} \\ \hline
(V) & Mean       & Std \%    & Mean      & Std \%     & Mean      & Std\%      \\ \hline
1.8 & 4.9$\times 10^{-6}$     & 1.22    & -2.0$\times 10^6$         &  $<$1          &   2.6$\times 10^{-6}$        &      3.74     \\ \hline
1.5 & 4.9$\times 10^{-6}$     & 1.25    & -1.26$\times 10^6$          &   $<$1         &   2.6$\times 10^{-6} $       &    3.65        \\ \hline
1.0 & 2.2$\times 10^{-6}$     & 3.81    & -1.9$\times 10^6$          &  $<$1         &    1.0$\times 10^{-6} $      &     5.68      \\ \hline

\end{tabular}%
}
\caption{The variation of the sigmoid function parameters (Monte Carlo Simulation with 200 samples) on scaling VDD}
\label{}
\end{table}
In Fig 9,(i) Transfer characteristic obtained from the circuit at Vdd = 1.8 (ii) Best Fit sigmoid function at Vdd = 1.8 (iii) Transfer characteristic obtained from the circuit at Vdd = 1.5 (iv) Best Fit sigmoid function at Vdd = 1.5 (i) Transfer characteristic obtained from the circuit at Vdd = 1.0 (ii) Best Fit sigmoid function at Vdd = 1.0. With Vdd = 1.0, we have decreased all the bias current in the circuit so that the transistors operate near saturation region.Current is mirrored by the factor of 1 in both the left and right current mirror so if we will increase the factor of the current mirror, it will further reduce the power.
\section{Conclusion}
In this paper, we have shown the degradation in the performance of RCM for dot product operation due to terminal resistance which is the input impedance of the neuron circuit. Error in the dot product operation increases with increase in the terminal resistance and decrease in the mem-resistance. To avoid this degradation, we have proposed current mode neuron which has low input impedance and more power efficient compared to the existing neuron. \newline
Power Consumption of Current Mode Neuron is half compared to voltage mode neuron at VDD = 1.8. Current mode neuron is much faster compared to voltage based neuron. On applying full input swing, the time required for the output to saturate is 4ns and 40ns in current and voltage mode respectively.Voltage mode has higher bandwidth compared to current mode neuron, but we can not operate at such high frequency because input impedance is large at that frequency and the equation (4) is no longer valid.The current mode neuron circuit does not require any large resistor, so it is more area efficient.The sigmoid transfer characteristic (Fig. 9) obtained from current mode neuron matches more perfectly with an ideal sigmoid function, while a deviation is observed in the voltage mode transfer characteristic (Fig. 5).From II and V, we observe that current mode neuron can be operated at lower VDD with less variation in parameters a ,b and c of the sigmoid function.

\bibliographystyle{plain}
\bibliography{ref.bib}

\end{document}